\documentclass[12pt,a4paper]{article}
\usepackage[utf8]{inputenc}
\usepackage[T1]{fontenc}
\usepackage{amsmath,amssymb,amsfonts}
\usepackage{graphicx}
\usepackage{geometry}
\usepackage[numbers,sort&compress]{natbib}
\usepackage{caption}
\usepackage{subcaption}
\usepackage{setspace}
\usepackage{authblk}
\usepackage{xcolor,xspace}
\usepackage{placeins}
\usepackage{booktabs}

\usepackage[colorlinks=true, linkcolor=blue, citecolor=blue, urlcolor=blue]{hyperref}
\usepackage{dblfloatfix}
\usepackage[export]{adjustbox}
\usepackage{tikz}
\usepackage{float}
\usepackage{comment}
\DeclareRobustCommand{\circnum}[1]{%
	\tikz[baseline=(char.base)]{
		\node[shape=circle,draw,inner sep=0.8pt] (char) {\footnotesize #1};
	}%
}
\graphicspath{{figures/}{sections/}}
\newcommand{\method}{\textbf{Methods and Materials}\xspace}
\newcommand{\rgA}{\textbf{Accommodating}\xspace}
\newcommand{\rgF}{\textbf{Friction-averse}\xspace}
\newcommand{\rgG}{\textbf{Guarded}\xspace}
\newcommand{\rgE}{\textbf{Evaluative}\xspace}
\newcommand{\ptA}{\textbf{A}\xspace}
\newcommand{\ptF}{\textbf{F}\xspace}
\newcommand{\ptG}{\textbf{G}\xspace}
\newcommand{\ptE}{\textbf{E}\xspace}
\newcommand{\ptB}{\textbf{B}\xspace}
\newcommand{\ptC}{\textbf{C}\xspace}
\newcommand{\ptD}{\textbf{D}\xspace}
\newcommand{\rgptA}{\textbf{Accommodating} (\textbf{A})\xspace}
\newcommand{\rgptF}{\textbf{Friction-averse} (\textbf{F})\xspace}
\newcommand{\rgptG}{\textbf{Guarded} (\textbf{G})\xspace}
\newcommand{\rgptE}{\textbf{Evaluative} (\textbf{E})\xspace}

\newcommand{\rrptF}{\textbf{Friction-averse} regime (\textbf{F})\xspace}

\newcommand{\rrptE}{\textbf{Evaluative} regime (\textbf{E})\xspace}
\newcommand{\nameER}{{Erd\H{o}s--R\'enyi}\xspace}
\newcommand{\nameBA}{{Barab\'asi--Albert}\xspace}
\newcommand{\methods}{{\bf Methods and Materials}\xspace}
\newcommand{\pl}[1]{\texttt{#1}\xspace}

\newcommand{\yrl}[1]{}
\newcommand{\rml}[1]{}
\newcommand{\rmd}[1]{}

\makeatletter
\newcommand{\GNCsilabel}[2]{%
  \expandafter\gdef\csname GNC@si@#1\endcsname{#2}}
\GNCsilabel{app:trend_1D_case}                        {S1.2}
\GNCsilabel{SI:belief-trajects-ps1}                   {S1.3}
\GNCsilabel{fig:trajectories_ps1}                     {S2}
\GNCsilabel{app:polarization-role-modular-synthetic}  {S1.4.1}
\GNCsilabel{SI:rho-role-FG}                           {S1.5}
\GNCsilabel{app:Empirical Networks-bitcoin}           {S1.6.1}
\GNCsilabel{app:legislative network}                  {S1.6.2}
\GNCsilabel{SI:credence-background}                   {S1.7}
\AtBeginDocument{%
  \let\GNC@origref\ref
  \renewcommand{\ref}[1]{%
    \@ifundefined{GNC@si@#1}%
      {\GNC@origref{#1}}%
      {\csname GNC@si@#1\endcsname}}%
}
\makeatother

\title{Regimes of Influence under Trust--Distrust Gating}
\author[1]{Razieh Masoumi}
\author[1]{Ahana Biswas}
\author[1]{Yu-Ru Lin}
\affil[1]{School of Computing and Information, University of Pittsburgh, Pittsburgh, PA, USA}
\date{\today}

\begin{document}
\maketitle

\begin{abstract}
In many social communities, individuals can simultaneously trust and distrust the same source, a feature standard opinion-dynamics models often ignore. We formalize this ambivalence with \textit{Gated Network Credence}, in which each directed relationship encodes distinct trust and distrust assessments. These jointly determine ``net trust''---the willingness to rely on a source---and ``uncertainty''---the conflict between trust and distrust within the same relationship. Agents update beliefs only when net trust exceeds a threshold and uncertainty falls below another, yielding an \textit{effective influence graph} whose topology drives long-run belief states. Sweeping both thresholds uncovers four regimes---Accommodating, Evaluative, Friction--averse, and Guarded---that differ in openness to trust and conflict. 
Under the modeled family of trust--distrust coupling and prestige-biased trust allocation, the model exhibits a hub--periphery reversal: in the Evaluative regime, high-degree agents contribute more strongly to the limiting belief, whereas in the Friction--averse regime, stringent uncertainty filtering can disproportionately remove hub-directed influence channels and reduce their contribution to the collective equilibrium. This conditional pattern recurs across synthetic network topologies and two empirical network evaluations.
Our results show that belief dynamics depend not only on network structure but also on how relational ambivalence between trust and distrust gates interpersonal influence.

\end{abstract}

\section{Introduction}
Understanding how individuals form, revise, and propagate opinions is essential for explaining collective phenomena such as polarization, echo chambers, and radicalization~\cite{noorazar2020, galesic2021, moscovici1969, sunstein2002, galam2002, castellano2009, iyengar2012, baumann2020}. Classical models of opinion dynamics such as De Groot's averaging scheme~\cite{degroot1974}, Hegselmann-Krause bounded-confidence (BC) updates~\cite{hegselmann2002, deffuant2000, bernardo2024}, or the Sznajd local-persuasion rule~\cite{sznajd2000}, capture a salient feature of human interaction: individuals revise their beliefs through social exposure, whether by averaging neighbors' states, interacting only within a confidence bound, or responding to local persuasion. Subsequent extensions have enriched these opinion-dynamics frameworks with mechanisms such as signed bounded confidence~\cite{altafini2018} and heterogeneous agent activity~\cite{li2023}. Existing models explain how consensus can arise, but they usually treat interpersonal influence as a single channel that is either active or inactive, positive or negative, or stronger or weaker, which makes it difficult to represent the possibility that the same source is simultaneously trusted and distrusted.

Such ambivalence is common in everyday life---one may respect an expert's credentials while remaining skeptical of their motives, or consider a news outlet reliable for some topics but unreliable for others. Empirical studies confirm that trust and distrust are distinct constructs that can coexist toward the same source, producing ambivalence in relationships and shaping how individuals perceive and recall information~\cite{lewicki1998, moody2014, moody2017, posten2021}. Because trust and distrust shape receptivity in different ways, distinguishing them is necessary for understanding when individuals revise their views and when they reject discordant information outright, with direct consequences for misinformation spread and the persistence of opinion clusters~\cite{lazer2018}. Yet most opinion-dynamics models collapse them into a single scalar or treat them as mutually exclusive signs. This binary representation is also reflected in widely used signed-network benchmarks: for instance, the Epinions and Slashdot networks encode each tie as either positive, indicating trust, or negative, indicating distrust, but not both~\cite{guha2004, kunegis2009, leskovec2010}. Signed-network models built on this representation often inherit the same mutual-exclusivity assumption in studies of trust prediction and consensus formation~\cite{kou2021, fang2022, zha2023, shi2020}. A few studies have attempted to incorporate trust and distrust as separate components of opinion dynamics~\cite{ishii2019gdn, ishii2019wi, ishii2020taai, taghavi2020}, but they typically map them onto opposite-valued influence parameters rather than representing trust and distrust as two independent relational attributes that can coexist within the same directed relationship. This simplification limits our ability to predict when a network will fragment into multiple opinion clusters or when strongly polarized or initially extreme belief states can spread.

To address this gap, we introduce \textit{Gated Network Credence}, an opinion-dynamics model that represents each directed relationship as a pair of relational evaluations: the trust an agent assigns to a source and the distrust the same agent assigns to that source. These two evaluations jointly determine \textit{net trust}---the willingness to rely on a source---and \textit{uncertainty}---the degree of trust--distrust conflict within the relationship. Influence is therefore gated at the relationship level: an agent incorporates a neighbor's belief only when the agent's ``net trust'' evaluation  of that neighbor is sufficiently positive and the trust--distrust conflict in that relationship is sufficiently low. The set of active directed relationships that survive these dual filters defines an \textit{effective influence graph}; the topology of this graph governs the trajectory and equilibrium of agents' belief states.

We study this gated updating mechanism by running simulations on three synthetic networks---\nameER, \nameBA, and modular networks---and on two real-world empirical networks: the Bitcoin-OTC reputation network and a directed follow network of U.S.\ state legislators on Twitter/X.

To isolate how the gate conditions the influence of structurally prominent sources, we conduct a controlled mechanism test in which the most {\it positive} sampled initial beliefs are assigned to a small set of highest-degree agents. This scenario is motivated by evidence that structural position affects diffusion and attention, and that prominent actors can play consequential roles in social communication and information spread \citep{kitsak2010,centola2010}. We use ``positive'' beliefs as a reference direction: because the simulated initial-belief distribution is symmetric around zero and the updating rule is invariant to a global reversal of belief signs, seeding the most negative beliefs would produce the corresponding mirror-image test. Using the positive direction as a reference also makes the interpretation direct: an increase in the limiting belief indicates stronger representation of the initially seeded direction, whereas a decrease indicates that its contribution is reduced. The design therefore allows us to test whether beliefs initially concentrated among prominent agents retain their contribution to the collective outcome or lose influence when their influence channels fail the trust--distrust gate.
 
By varying how permissive agents are toward net trust and trust--distrust conflict, we identify four regimes (Figure~\ref{fig:phase-diagram}(a)\textcircled{3}): \rgA, \rgE, \rgF, and \rgG. These regimes produce qualitatively different collective belief outcomes.
We label these conditions according to the relational admission rule induced by the two thresholds. The \rgA regime admits a broad range of sources because it is permissive toward both net trust and relational conflict. The \rgE regime evaluates sources primarily through a more selective net-trust requirement while tolerating conflict. The \rgF regime admits sources with sufficient net trust while avoiding relationships marked by elevated trust--distrust conflict. The \rgG regime applies both requirements restrictively, thereby admitting only relationships that are strongly trusted and low in conflict.

Under the specified trust--distrust coupling, prestige allocation, and degree-biased seeding procedure, the \rgE and \rgF regimes can exhibit a \textit{hub--periphery reversal}. In the \rgE regime, high-degree seeded agents retain greater direct influence on the limiting belief because more hub-directed relationships survive the gate. Conversely, in the \rgF regime, the conflict filter can remove many hub-directed relationships and reduce the direct contribution of positively seeded hubs. When the same initial extreme beliefs are placed on hubs, the collective belief can therefore move away from the seeded extreme position. The same qualitative contrast appears in the two empirical network evaluations, indicating that the result is not limited to a single synthetic topology.

These findings suggest that analyses of interventions should distinguish whether an intervention is expected to change relational evaluations, shared admission criteria, or the network of available exposure. By formalizing influence as a gated process, in which interpersonal ties become active only under specific configurations of net trust and trust--distrust conflict, our framework offers a lens for reasoning about the structural conditions under which seeded extreme positions spread or fail to take hold. 
\section{Results}

\subsection{Trust--distrust ambivalence yields qualitatively distinct regimes of collective belief dynamics}\label{sec:2.1}

\begin{figure*}[!t]
    \centering
    \includegraphics[width=.98\linewidth]{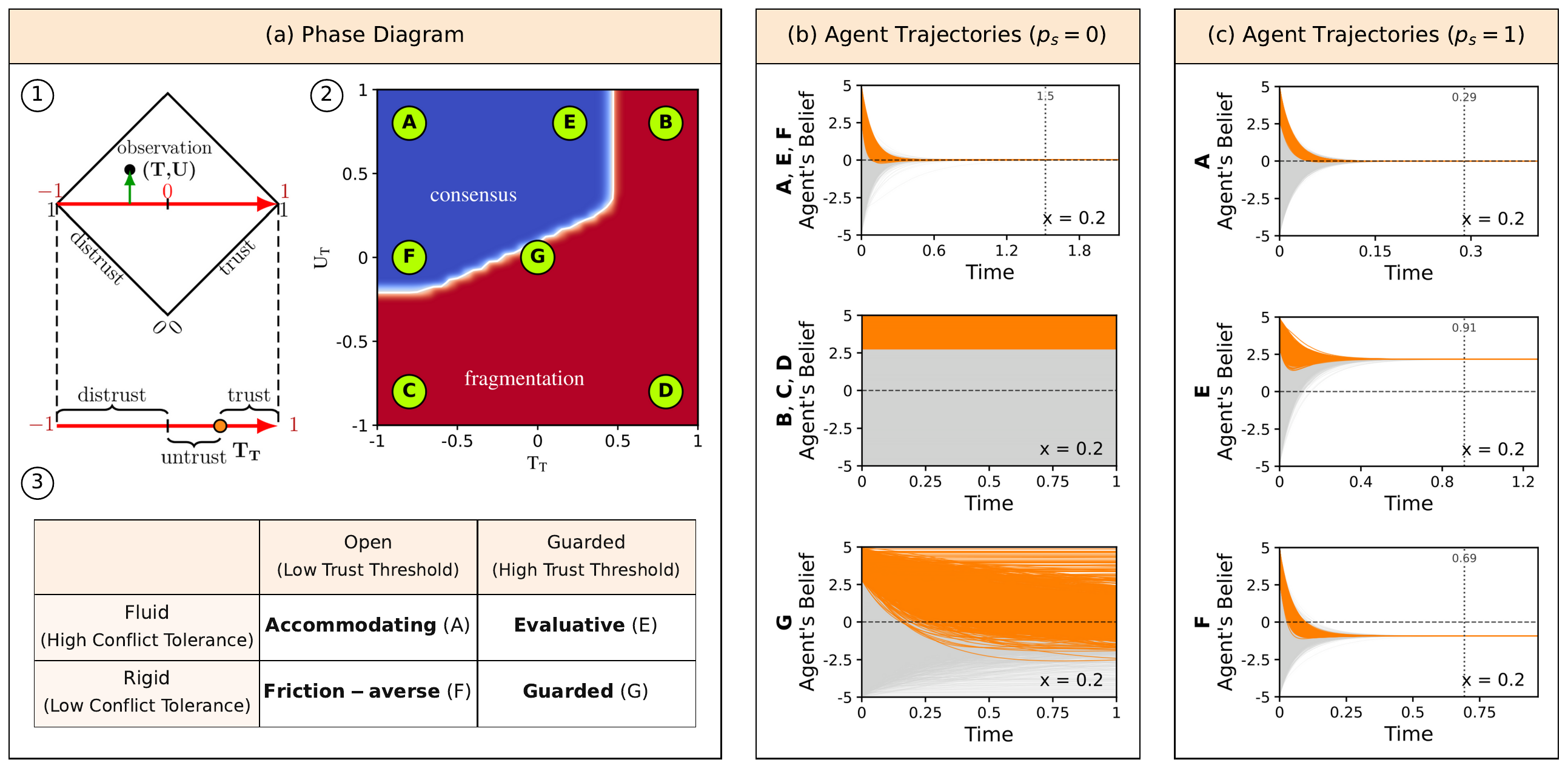}
    \vspace{1em}
\caption{
{\bf (a)}\textcircled{1} Two-dimensional credence space for a directed relationship with separate trust $\tau$ and distrust $\delta$ assessments. The coordinate $(T,U)$, indicating net trust and uncertainty, is obtained from $(\tau,\delta)$ via $T=\tau-\delta$ and $U=\tau+\delta-1$. The horizontal line $U=0$ corresponds to the conventional one-dimensional case in which trust and distrust are treated as perfectly coupled opposites.
\textcircled{2} Phase diagram of long-run belief dynamics on an \nameER graph $(N=5000,\ p=0.01)$ under fixed simulated conditions. Blue indicates global consensus, red indicates persistent fragmentation, and the white curve marks the phase boundary.
\textcircled{3} The two gating thresholds partition the phase diagram into four regimes: agents may be more or less permissive toward net trust, and more or less tolerant of trust--distrust uncertainty. This yields four distinct regimes---\rgptA, \rgptE, \rgptF, and \rgptG.
{\bf (b)} Belief trajectories for thresholds under prestige bias $p_s=0$; orange curves indicate radical promoters, gray curves indicate other agents. In panels that reach consensus, the vertical dotted line indicates the convergence time $T_{\mathrm{final}}$, annotated above, defined as the first time at which the belief dynamics settle into their stationary state. The subplots for points \ptA, \ptE, and \ptF in the consensus region show that the system converges to a moderate final belief state centered near zero.
{\bf (c)} Same conditions as (b) but with maximal prestige bias \(p_s=1\). The \rgptE regime now produces consensus aligned with the positive radical seeds, whereas the \rgptF regime leads to an opposite-signed stationary consensus. Throughout, the trust--distrust coupling is fixed at $\rho=-0.4$. See \methods for study methodology.
}
    \label{fig:phase-diagram}
\end{figure*}

We show that representing trust and distrust as distinct relational assessments within the same social relationship, rather than as opposite ends of a single continuum, allows the dual-threshold gate to produce qualitatively distinct regimes of interpersonal influence.

We represent agents as nodes and directed edges as interpersonal relationships. Each directed relationship $(i, j)$ carries a trust assessment $\tau_{ij} \in [0,1]$ and a distrust assessment $\delta_{ij} \in [0,1]$; the procedure for generating $\tau_{ij}$ and $\delta_{ij}$, including their coupling structure, is described in \method~\ref{app:trust-distrust-coupling-definition}. We then map each pair of trust and distrust via:
\[
(T_{ij}, U_{ij}) = (\tau_{ij} - \delta_{ij}, \tau_{ij} + \delta_{ij} - 1),
\]
which defines a two-dimensional credence space : $T_{ij}$ captures the net inclination of agent $i$ toward $j$, whereas $U_{ij}$ is an operational coordinate of trust--distrust conflict (Figure~\ref{fig:phase-diagram}(a)\circnum{1}). The value of $U_{ij}$ increases when trust and distrust are jointly elevated relative to the baseline $U=0$ and decreases when both evaluations are jointly weak. We use the term ``uncertainty'' as shorthand for this form of relational conflict. This construction is motivated by theories of evaluative space and relational ambivalence, which treat positive and negative evaluations as distinguishable dimensions rather than opposite poles \cite{cacioppo1997beyond,lewicki1998,moody2014}. 

The conventional one-dimensional notion of trust is recovered when trust and distrust are treated as perfectly coupled opposites, corresponding to the horizontal line $U=0$. Away from this line, two relationships can have the same net trust $T$ but different uncertainty $U$. For example, high trust combined with high distrust and low trust combined with low distrust can produce the same net inclination while representing different degrees of relational ambivalence.
Thus, the additional coordinate changes which relationships are admitted by thresholding. In a one-dimensional model, raising the trust threshold can only remove relationships from the active set: every relationship that remains active at a stricter threshold would also have been active at a more permissive threshold. The dual-threshold rule does not impose this single ordering. One relationship may have higher net trust but higher uncertainty than another, so changing $(T_T,U_T)$ can retain either relationship, both relationships, or neither relationship. The resulting effective influence graphs are therefore not generally obtainable by thresholding a single trust score.

Under \textit{Gated Network Credence}, influence from $j$ to $i$ occurs only when agent $i$ has sufficiently high net trust in $j$ and the trust--distrust uncertainty associated with that relationship is sufficiently low:
\begin{equation}\label{eq:decision}
T_{ij} \ge T_T \quad \text{and} \quad U_{ij} \le U_T .
\end{equation}
Here, $T_T$ is the minimum net trust threshold required for a belief update, and $U_T$ is the maximum uncertainty threshold an agent is willing to tolerate; in simulations, both thresholds are swept over $[-1,1]$. 
Thus, a source can be rejected either because net trust is insufficient or because trust--distrust conflict is too high.
Further conceptual background and connections to established trust typologies are provided in Appendix~\ref{SI:credence-background}.

The initial belief of agent $i$, $b_i(0)$, was drawn uniformly from $[-B_{\max},\, B_{\max}]$. To assess how network structure and gating jointly shape collective belief outcomes, we conduct a simulated experiment in which the most positive initial beliefs are exogenously assigned to the highest-degree agents. We vary four empirically motivated parameters: {\it promoter prevalence}, $x$, controls how widely these positive initial beliefs are distributed among highly connected agents; {\it prestige bias}, $p_s$, controls the degree to which trust is aligned with source prominence; {\it trust--distrust coupling}, $\rho$, controls how strongly trust and distrust are inversely coupled within a relationship; and, in modular networks, {\it polarization}, $\psi$, controls the initial separation of beliefs between the two communities. Together, these parameters allow us to examine how prominence, relational conflict, and community separation interact with the gate. Full simulation details are provided in  \method section~\ref{sec:simulation-setup}.

\paragraph{Consensus–-fragmentation phase transition}

Figure~\ref{fig:phase-diagram}(a)\textcircled{2} shows the long-run outcome under \textit{Gated Network Credence} across the two-dimensional threshold space defined by $(T_T,U_T)$ for an \nameER network. Blue shading indicates threshold values that drive all agents to a single final belief state (global consensus), while red shading covers thresholds where the effective influence graph fragments and opinion clusters persist indefinitely. The white curve delineates the phase boundary between these two phases.

In the blue region, the threshold pair $(T_T, U_T)$ leaves enough links simultaneously satisfying both $T_{ij} \geq T_T$ and $U_{ij} \leq U_T$ to sustain a connected effective influence graph that drives the network to global consensus. At point $\mathbf{A}$ (permissive $T_T$, permissive $U_T$), both thresholds are highly permissive, so the vast majority of links meet both conditions, yielding a dense effective graph. We accordingly use point $\mathbf{A}$ as the reference point, assessing the permissiveness or restrictiveness of every other point relative to $\mathbf{A}$. At point $\mathbf{E}$ (more restrictive $T_T$, permissive $U_T$), the trust threshold is selective: only relationships with sufficiently high net trust remain active. Because the uncertainty threshold is permissive, it removes few additional relationships, so the effective graph is sparser but still connected. At point $\mathbf{F}$ (permissive $T_T$, more restrictive $U_T$), the uncertainty threshold is selective: highly conflicted relationships are removed, while the permissive trust threshold allows most low-uncertainty relationships to remain active. 

In contrast, the red region corresponds to threshold values where the effective influence graph becomes disconnected because too few relationships survive the dual gate. In this case, belief updating proceeds only within disconnected components, producing persistent belief clusters. The number of such clusters grows when either $T_T$ or $U_T$ becomes more restrictive: as fewer relationships remain active, the effective graph breaks into progressively smaller components. A very low $U_T$ as at point \ptC removes most relationships that carry even moderate uncertainty, whereas a high $T_T$ as at points \ptB and \ptD requires very strong net trust before a belief update can occur. Point \ptG($T_T=0$, $U_T=0$) lies near the consensus--fragmentation boundary, where both filters are restrictive. Because the resulting effective graph is sparse, small changes in parameters can determine whether the system reaches large-scale consensus or separates into fragmented belief clusters.

The \rgE and \rgF regimes are especially informative because each makes one admission criterion selective while leaving the other relatively permissive. In the \rgE regime, the net-trust threshold is selective and the conflict threshold is permissive; in the \rgF regime, the reverse holds. 
Comparing these conditions therefore highlights how the trust--distrust conflict dimension, absent from conventional one-dimensional trust models, changes collective belief outcomes. 
We next examine how degree-biased placement of positive extreme beliefs interacts with these two filtering logics.

\paragraph{Prestige bias differentiates collective belief outcomes in the \rgE and \rgF regimes.}
Figure~\ref{fig:phase-diagram}(b) first shows belief trajectories in the absence of prestige bias ($p_s=0$). In this baseline, trust--distrust pairs are assigned independently of source degree. Although the most positive initial beliefs are seeded on hubs, both the \rgE and \rgF regimes converge to a moderate collective equilibrium near $b^* \approx 0$. 
Thus, degree-biased placement of positive extreme beliefs alone does not visibly shift the collective equilibrium when trust is unrelated to source prominence.

The trajectories differ when prestige bias is introduced. As $p_s$ increases, agents assign greater trust to well-connected neighbors, making hub-directed relationships more likely to remain in the effective influence graph. Because the seeding procedure concentrates the most positive initial beliefs among high-degree nodes while lower-degree nodes remain negative on average (Section~\ref{subsec:promoter-prevalence-mech}), the effect of prestige bias depends on how each regime filters hub-directed relationships.
Figure~\ref{fig:phase-diagram}(c) illustrates this divergence under maximal prestige bias ($p_s=1$). In the \rgE regime, the stationary belief shifts toward the positive seeded beliefs, and this shift becomes larger as the fraction of promoters increases (Figures~\ref{fig:trajectories_ps1}(e)--(h) in Appendix~\ref{SI:belief-trajects-ps1}). In the \rgF regime, the equilibrium instead shifts away from the positive seeds and converges to an opposite-signed consensus (Figures~\ref{fig:trajectories_ps1}(i)--(l) in Appendix~\ref{SI:belief-trajects-ps1}). The same degree-biased seeding and prestige-aligned trust produce opposite collective outcomes.

\begin{figure*}[t]
    \centering
    \includegraphics[width=.6\textwidth]{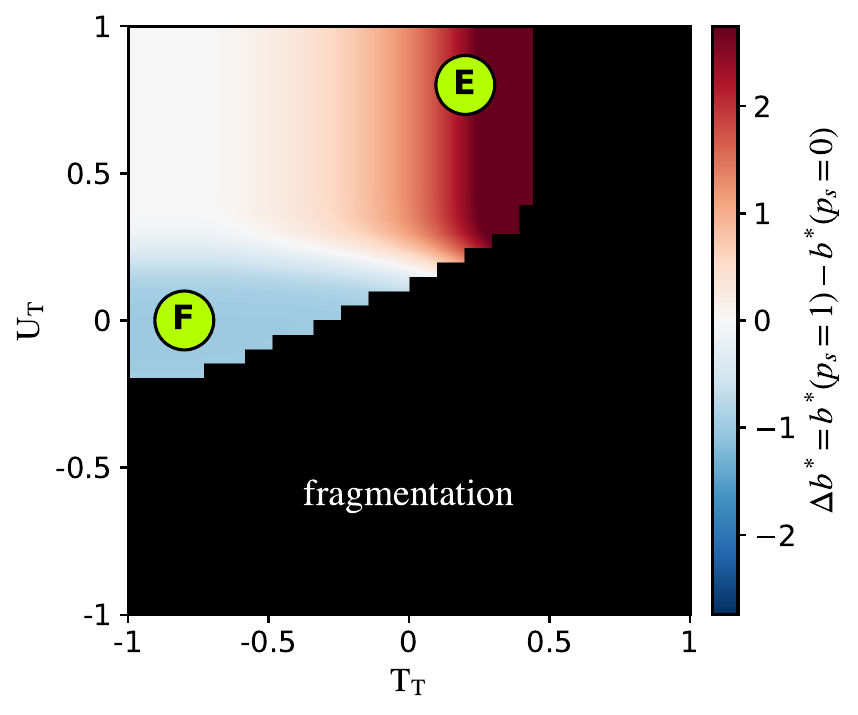}
\caption{\textbf{The sign reversal occupies extended threshold regions.} Change in the long-run collective belief under prestige bias, $\Delta b^* = b^*(p_s{=}1)-b^*(p_s{=}0)$, across the threshold plane $(T_T,U_T)$ for an \nameER network ($N=5000$, $p=0.01$, $\rho=-0.4$). Red indicates a shift toward the positive seeded beliefs, blue indicates a shift away from them, and black denotes fragmentation. Points $\mathbf{E}$ and $\mathbf{F}$ are representative settings in the positive and negative regions associated with the \rgE and \rgF regimes, respectively.}
  \label{fig:sign-reversal-robustness}
\end{figure*}

Figure~\ref{fig:sign-reversal-robustness} maps
$\Delta b^*=b^*(p_s=1)-b^*(p_s=0)$
over the consensus region of the threshold plane $(T_T,U_T)$. The black region denotes fragmentation, for which $\Delta b^*$ is not defined. Within the consensus region, the response contains three qualitatively distinct areas. In the red region, $\Delta b^*>0$, so increasing prestige bias shifts the equilibrium toward the positive seeded beliefs. The largest positive values occur near the phase boundary at high $T_T$. In the blue region, $\Delta b^*<0$, so increasing prestige bias shifts the equilibrium away from the positive seeded beliefs. A near-zero region separates these positive and negative responses, with the transition becoming narrower near the phase boundary.
The representative settings \ptE and \ptF lie within the positive and negative regions, respectively. The corresponding trajectories in Figure~\ref{fig:phase-diagram}(c) therefore illustrate the typical response of their surrounding threshold regions rather than behavior unique to two isolated parameter choices.

\begin{figure*}[t]
    \centering
    \includegraphics[width=1\textwidth]{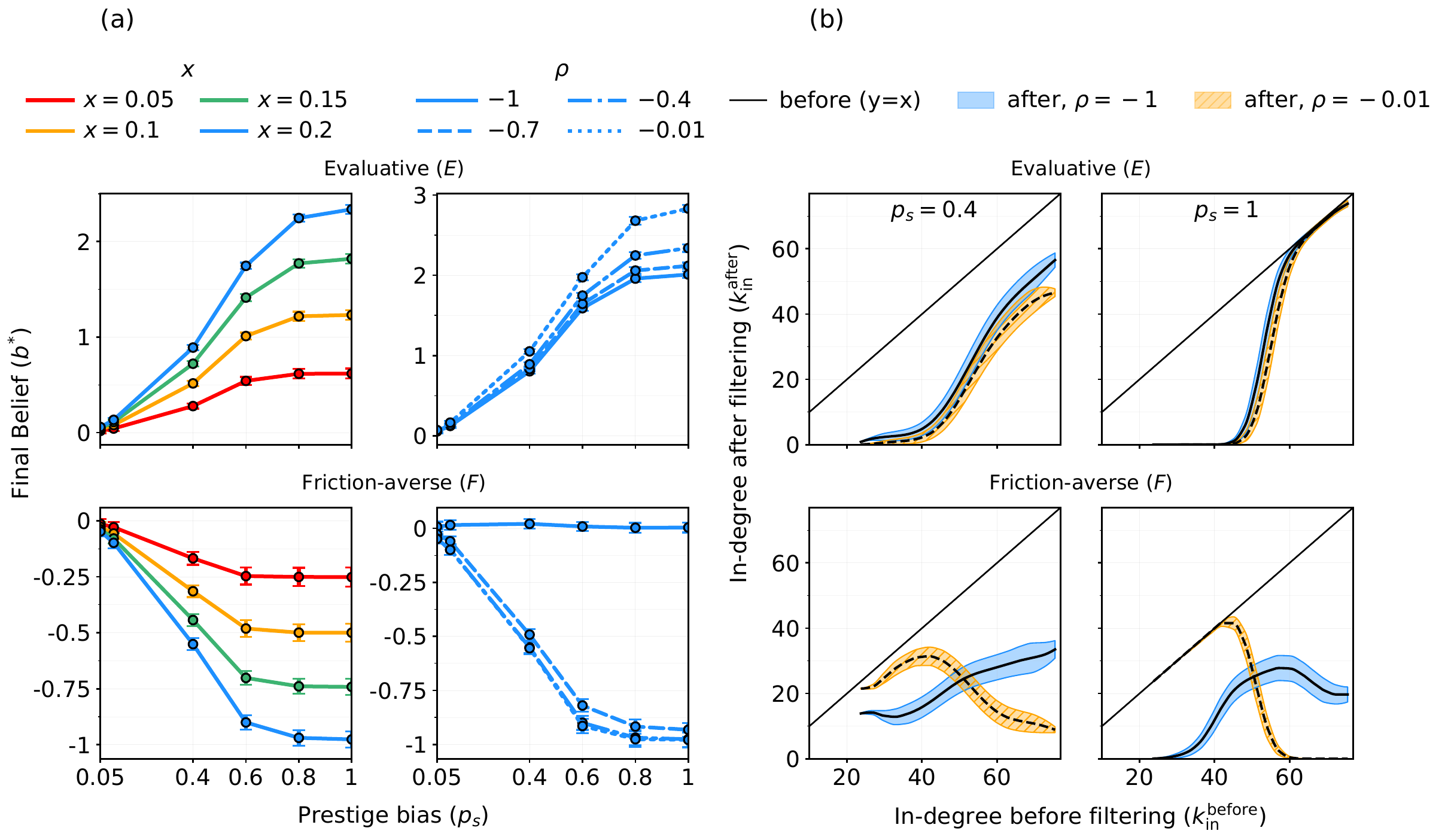}
\caption{
{\bf (a)} Long-run consensus belief state $b^{*}$ versus prestige bias $p_{s}$ on an \nameER network ($N=5000, p=0.01$). \pl{Top}: \rrptE; \pl{bottom}: \rrptF. \pl{Left} column fixes $\rho=-0.4$ and varies promoter prevalence $x$. \pl{Right} column fixes $x=0.2$ and varies trust--distrust coupling strength $\rho$. Error bars indicate the mean $\pm 1$ standard deviation across 100 independent realizations.
{\bf(b)} Mean and dispersion of post-filtering in-degree $k_{\mathrm{in}}^{\text{after}}$ as a function of pre-filtering in-degree $k_{\mathrm{in}}^{\text{before}}$, on the same \nameER network as in \pl{(a)}, single realization. \pl{Top} row: \rrptE; \pl{bottom}: \rrptF.
\pl{Left} and \pl{right} columns correspond to prestige bias values $p_s=0.4$ and $p_s=1$, respectively. Ribbons show the mean $\pm$ one standard deviation (SD) of post-filtering in-degree within each pre-filtering in-degree bin, computed across agents within the bin. Solid and dashed lines mark the corresponding mean post-filtering in-degree for $\rho=-1$ and $\rho=-0.01$, respectively; the black diagonal marks $k_{\mathrm{in}}^{\text{after}} = k_{\mathrm{in}}^{\text{before}}$ (no filtering). Mean lines and ribbon boundaries are smoothed across bins for display.
}
\label{fig:Trends-FG-Erdos}
\end{figure*}

\begin{figure}[!t]
\centering
\includegraphics[width=\linewidth]{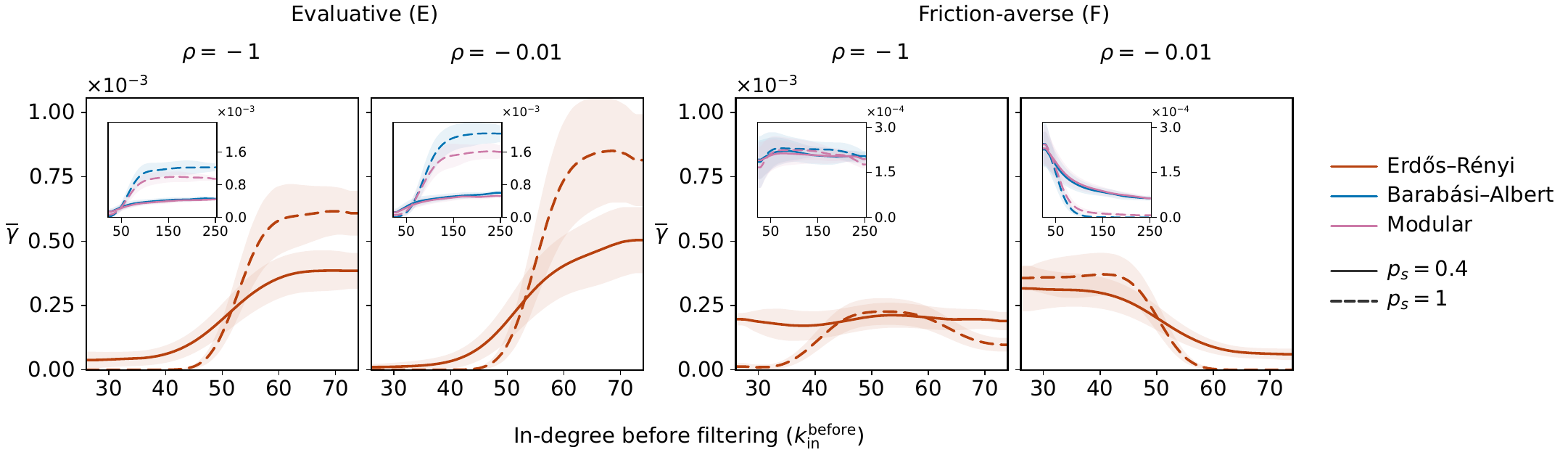}
\caption{%
Spectral characterization of long-run influence after filtering.
The normalized left zero eigenvector $\bar{\gamma}$ of the directed Laplacian of the filtered influence graph is plotted as a function of in-degree before filtering, $k_{\mathrm{in}}^{\mathrm{before}}$. Each entry $\bar{\gamma}_i$ measures how much agent $i$'s initial belief contributes to the long-run consensus or component-level equilibrium; larger values indicate greater long-run influence.
Main panels show results for \nameER (ER) networks
($N=5000$, $p=0.01$); 
insets show the corresponding results for \nameBA (BA) and modular networks on separate horizontal scales to accommodate their more skewed degree distributions. The two blocks distinguish the \rgptE regime (left) from the \rgptF regime (right), and within each block the two panels distinguish tightly coupled trust--distrust assessments $(\rho=-1)$ from weakly coupled assessments $(\rho=-0.01)$. Line color identifies network topology, and line style identifies prestige bias ($p_s=0.4$ solid; $p_s=1$ dashed). Within each pre-filtering in-degree bin, curves show the mean of $\bar{\gamma}_i$ across agents in that bin; shaded bands show one standard deviation around the bin mean. The figure shows that \rgE filtering concentrates long-run influence on higher-degree agents, whereas \rgF filtering flattens or reverses this degree--influence relationship, especially under weak trust--distrust coupling and stronger prestige bias.
	\label{fig:zero_left_eigenvec_FG_all_nets}
}
\end{figure}

\subsection{Mechanisms of regime-dependent belief reversal}%

Figure~\ref{fig:Trends-FG-Erdos}(a) displays the equilibrium collective belief state $b^{*}$ as a function of prestige bias $p_{s}$ across two regimes: the \rrptE on the top row, and the \rrptF on the bottom. The left column keeps the trust--distrust coupling $\rho$ fixed while varying promoter prevalence $x$, whereas the right column fixes $x$ and varies $\rho$.

In the \rrptE, the equilibrium response profiles increase with prestige bias $p_s$: stronger prestige bias shifts the final collective belief state in the positive direction. The magnitude of this shift depends on promoter prevalence, because larger values of $x$ place more positive initial beliefs on high-degree agents (Figure~\ref{fig:Trends-FG-Erdos}(a)~\pl{Top-left}). When trust and distrust are tightly inversely coupled ($\rho \rightarrow -1$), high trust is generally accompanied by low distrust. A broader range of relationships can therefore satisfy the net-trust threshold, and the surviving relationships are less strongly concentrated on hubs. As the coupling weakens and $\rho$ approaches zero, high trust is less reliably accompanied by low distrust. The net-trust filter then admits a more selective subset of high-trust relationships. Under prestige bias, these retained relationships are more often directed toward hubs, increasing the contribution of positively seeded high-degree agents to $b^*$ (see Appendix~\ref{SI:rho-role-FG} for the derivation).

By contrast, under the \rgF regime (F), equilibrium belief generally decreases as prestige bias $p_s$ increases. Except when trust and distrust are nearly perfectly inversely coupled ($\rho \approx -1$), the equilibrium belief is negative across the examined $(p_s, x)$ combinations, and its magnitude increases with prestige bias (Figure~\ref{fig:Trends-FG-Erdos}(a)~\pl{bottom row}).
This pattern depends on the trust--distrust coupling. When $\rho \approx -1$, trust and distrust behave almost as exact opposites, so $U_{ij}$ remains close to zero and is only weakly related to $\tau_{ij}$. The uncertainty gate is therefore only weakly responsive to the trust differences introduced by prestige bias, and the equilibrium remains close to zero as $p_s$ increases.  
As $\rho$ approaches 0, the inverse coupling between trust and distrust weakens, so uncertainty $U_{ij}$ increases more strongly with trust $\tau_{ij}$. Under prestige bias, relationships directed toward more prominent sources receive higher trust values and tend to have larger uncertainty values. These hub-directed relationships are consequently more likely to fail the \rgF uncertainty threshold.
The resulting reversal is therefore conditional on the modeled coupling between source prominence, trust, and trust--distrust conflict.

These results indicate that the two gating regimes are associated with opposite responses to the same degree-biased seeding and prestige-aligned trust: \rgE filtering shifts the collective belief toward the positive seeds, whereas \rgF filtering shifts it in the opposite direction.

We next separate two mechanisms behind this reversal: which ties survive the gate, and how much long-run influence the remaining connected agents have.

\paragraph{Filtering differentially retains hub- and peripheral-directed pathways.}

Because positive extreme beliefs are seeded on the highest-degree agents, the reversal under \rgF filtering suggests that conflict-sensitive filtering reduces the incoming pathways through which hubs can affect long-run belief outcomes.
Figure~\ref{fig:Trends-FG-Erdos}(b) shows how many incoming relationships each agent retains after filtering, plotted against that agent's original in-degree before filtering. The columns compare moderate and high prestige bias ($p_s=0.4$ and $p_s=1$), while the line styles compare tightly coupled and weakly coupled trust--distrust assessments ($\rho=-1$ and $\rho=-0.01$). Curves show the mean post-filtering in-degree within each pre-filtering in-degree bin, and ribbons show one standard deviation around that mean. The \pl{top} row corresponds to the \rgE regime and the \pl{bottom} to the \rgF regime.

In the \rgptE regime, mean post-filtering in-degree increases sharply with pre-filtering in-degree: lower-degree agents retain few incoming relationships, whereas higher-degree agents preserve many of their incoming links. This pattern becomes steeper as prestige bias increases, showing that prestige-biased trust makes relationships directed toward well-connected agents more likely to survive the net-trust filter.

The \rgF regime shows a different retention pattern because uncertainty, rather than net trust, is the main constraint. For $\rho=-1$, the mean post-filtering in-degree is nearly degree-independent, indicating that filtering does not strongly distinguish between low- and high-degree agents. For $\rho=-0.01$, the mean rises at low-to-intermediate degrees but declines for high-degree agents, showing that many hub-directed relationships are removed 
(Figure~\ref{fig:Trends-FG-Erdos}(b)~\pl{Bottom}). This difference is driven by how coupling shapes uncertainty. Under strong inverse coupling ($\rho=-1$), high trust is paired with low distrust, so even hub-directed relationships remain relatively unconflicted and the uncertainty filter removes relationships without a strong degree bias. Under weak coupling ($\rho=-0.01$), distrust does not fall as reliably when trust is high; therefore, prestige-biased relationships directed toward hubs can become highly uncertain and are selectively removed by the \rgF filter. The effect is most pronounced with stronger prestige bias 
(Figure~\ref{fig:Trends-FG-Erdos}(b)~\pl{Bottom-right}), where high-degree agents retain few incoming links after filtering. 
Thus, the two regimes reshape the pathways through which influence can propagate: \rgE filtering preserves more hub-directed relationships, whereas \rgF filtering can equalize link retention across degree classes or reduce the retained connectivity of hubs.

\paragraph{Spectral analysis identifies agents' direct long-run contributions.}

The retention patterns in Figure~\ref{fig:Trends-FG-Erdos}(b) show which ties survive filtering; the spectral analysis in Figure~\ref{fig:zero_left_eigenvec_FG_all_nets} shows how the filtered graph distributes agents' direct long-run contributions. 
 
As shown in the \methods, the normalized left null vector $\bar{\gamma}$ of the filtered graph's directed Laplacian identifies how each agent's initial belief enters the limiting belief. In a consensus graph,
the $\bar{\gamma}_i$ indicates the direct contribution of agent $i$'s initial belief to the shared limiting belief. In a fragmented graph, the corresponding reach-specific vector describes each agent's direct contribution to the limiting belief within that reach.
The number of agents that receive that limiting belief depends separately on the reach's size and accessibility. Larger coefficients therefore indicate greater direct long-run contribution.
Figure~\ref{fig:zero_left_eigenvec_FG_all_nets} plots agents' long-run influence against their raw pre-filtering in-degree for \nameER networks (main panels), with \nameBA and modular networks shown in the insets on a separate horizontal scale reflecting their more skewed degree distributions. The panels compare lower and higher prestige bias under tighter and weaker trust--distrust coupling.

In the \rgptE regime (\pl{top} row), asymptotic influence increases with pre-filtering in-degree: higher-degree agents have a larger effect on the eventual belief state than lower-degree agents. At low prestige bias, low-degree agents retain small but strictly positive asymptotic weight, remaining within the set of influential sources. This increasing pattern becomes stronger when prestige bias is higher and trust--distrust coupling is weaker, because the filtering process retains more relationships directed toward well-connected agents, while the direct long-run contributions of the lowest-degree agents approach zero. The insets show the same qualitative increase for \nameBA and modular networks, although the shape of the increase differs because these networks have more heterogeneous degree distributions. Thus, in the \rgE regime, higher-degree agents remain the main determinants of the long-run collective belief state.

In the \rgptF regime (\pl{bottom} row), the pattern changes because the uncertainty filter removes many conflicted relationships directed toward hubs. 
Under tighter trust--distrust coupling, direct long-run contributions vary little by degree, indicating that high-degree agents no longer dominate the equilibrium. Under weaker coupling, high-degree agents contribute less directly to the limiting belief than low- or intermediate-degree agents; under stronger prestige bias, their contribution can approach zero. The insets show the same qualitative shift for \nameBA and modular networks. These results help explain the belief reversal reported above: in the \rgE regime, positively seeded hubs exhibit larger direct long-run contributions, whereas in the \rgF regime those contributions are reduced and relatively greater contributions can remain among lower-degree agents. The reversal reflects how trust- and conflict-based filtering changes the set of sources that contribute directly to limiting beliefs across the tested network topologies.

To assess whether the reversal depends on the conflict coordinate rather than on selective net-trust filtering alone, Appendix~\ref{app:trend_1D_case} reports a matched one-dimensional comparison. At the four tested net-trust thresholds, no reversal is observed when the uncertainty coordinate is removed and influence is filtered only by $T_{ij}$. This comparison indicates that the additional conflict coordinate is required to produce the observed sign reversal.

\subsection{Regime-dependent hub--periphery reversal in empirical networks}

\begin{figure*}[t]
    \centering
    \includegraphics[width=0.85\textwidth]{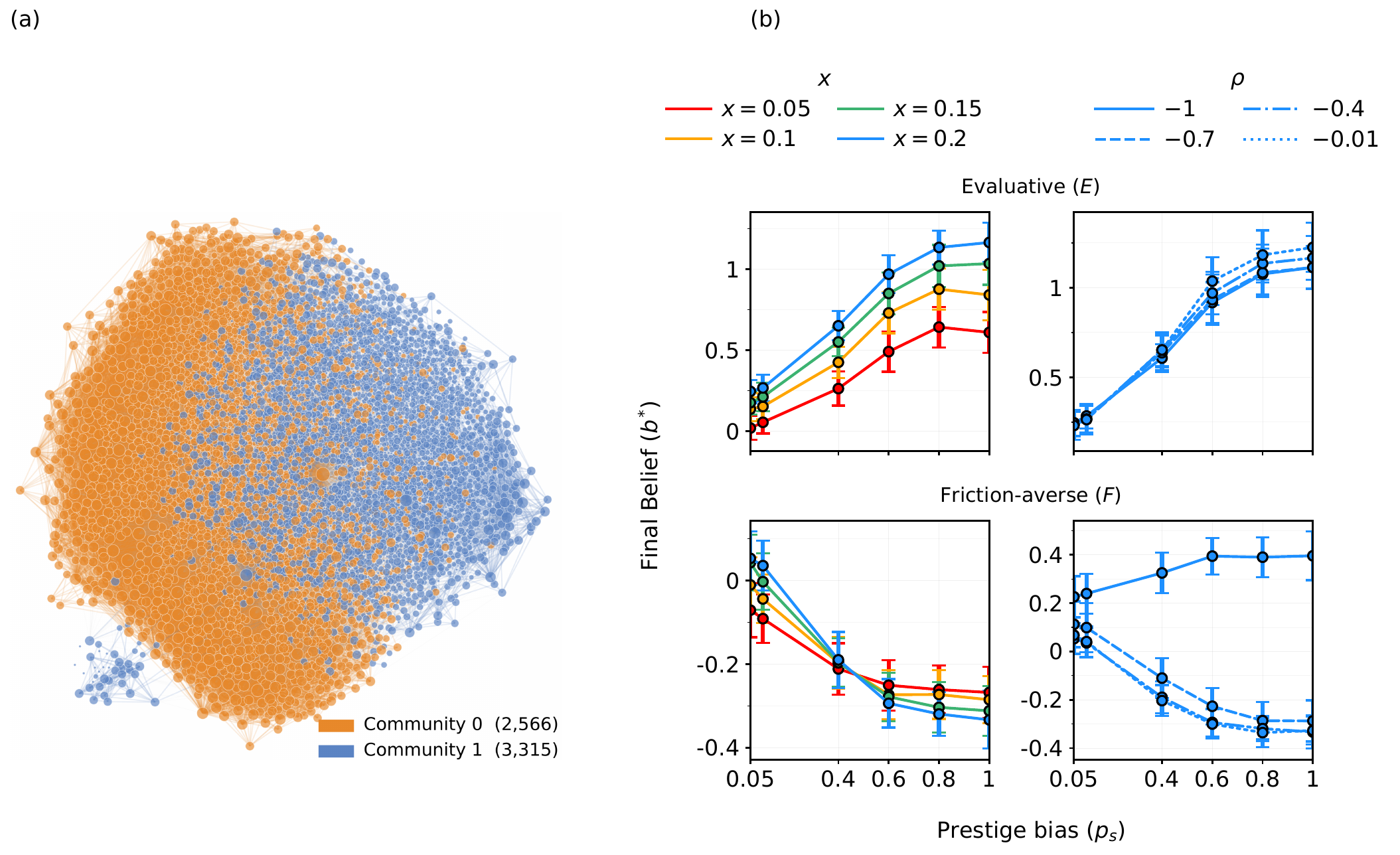}
    \caption{%
    {\bf (a)} Bitcoin-OTC reputation network; node color indicates community membership, and the legend reports the community sizes.
    {\bf (b)} Network-averaged long-run belief $\langle b^\ast\rangle$ versus prestige bias $p_s$ for the \rrptE
    in the \pl{top} row and the \rrptF in the \pl{bottom} row.
    \pl{Left} panels fix trust--distrust coupling $\rho=-0.4$; lines correspond to different positive promoter fractions $x$.
    \pl{Right} panels fix $x=0.2$; line styles represent different coupling strengths $\rho$.
    Error bars represent one standard deviation across 100 independent runs with randomized initial beliefs. }
    \label{fig:bitcoin-otc-with-network}
\end{figure*}

\begin{figure*}[!t]
    \centering
    \includegraphics[width=\textwidth]{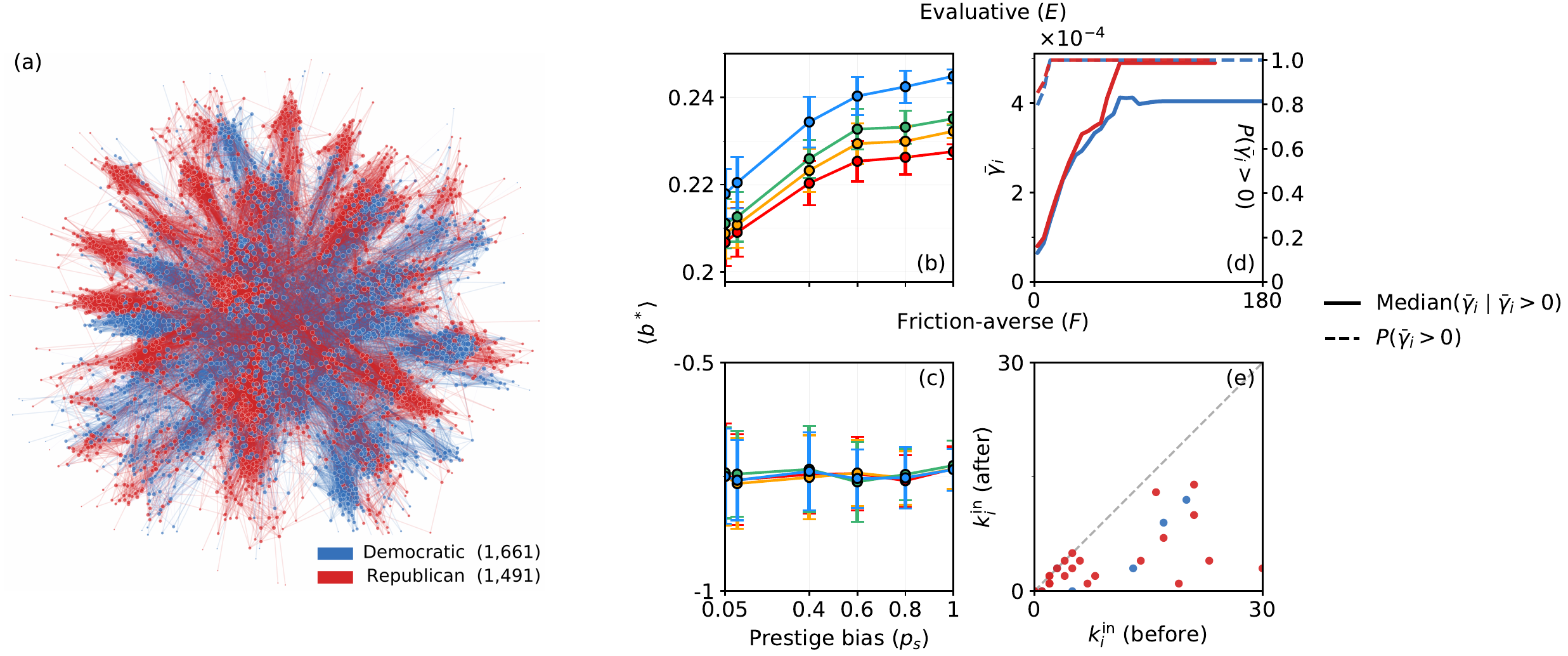}
\caption{%
{\bf (a)} Directed follow network of U.S. state legislators on X/Twitter: node color encodes legislators' party affiliation (blue for Democrats, red for Republicans).  
{\bf (b-c)} Network-averaged long-run belief $\langle b^\ast\rangle$ versus prestige bias $p_s$ under the \rgptE and \rgptF regimes. Lines are color coded by positive promoter fraction $x \in \{0.05, 0.10, 0.15, 0.20\}$, using the same color scheme as Fig.~\ref{fig:bitcoin-otc-with-network}(b). 
{\bf (d)} Binned summaries of the long-run influence vector $\bar{\gamma}$ plotted against pre-filtering in-degree for both communities under the \rgE regime.  
Dashed and solid lines represent, respectively, the fraction of legislators with positive long-run influence, $P(\bar{\gamma}_i>0)$, and the median long-run influence conditional on being positive, $\mathrm{Median}(\bar{\gamma}_i \mid \bar{\gamma}_i>0)$. Both quantities are computed within pre-filtering in-degree bins and smoothed across bins.
{\bf (e)} Pre- and post-filtering in-degree for legislators in isolated strongly connected components under the \rgF regime; the distribution is skewed toward low-degree Republican legislators. Panels {\bf (d)} and {\bf (e)} use $p_s = 1$.}
\label{fig:trend-legislative-with-network}
\end{figure*}

We next evaluate whether the hub--periphery reversal observed in synthetic networks also appears when the model is instantiated on real-world networks with empirical relational information. The Bitcoin-OTC reputation network provides observed graded trust ratings, while distrust is generated through the controlled coupling model. The U.S. state-legislator follow network provides two relational inputs constructed from distinct observed structural and ideological signals. The two cases differ in the extent to which trust and distrust are empirically grounded.

\paragraph{Evidence from the Bitcoin--OTC reputation network.}
The Bitcoin-OTC graph consists of \(\textbf{5{,}881}\) users and \(\textbf{35{,}592}\) directed rating ties distributed across two communities (Figure~\ref{fig:bitcoin-otc-with-network}(a)). We construct trust assessments from observed user-to-user rating scores. For each directed rating, the score is rescaled to the unit interval and used as the mean of the edge-specific trust distribution; distrust is then generated from the sampled trust value through the coupling as in the synthetic analyses (see \methods, Eq.~\ref{eq:trust_distrust_coupling}).
Because the two empirical communities differ in size, we use the larger community as the focal seeded community and apply the same within-community initialization rule used in the synthetic analyses. This preserves a consistent promoter-prevalence and degree-biased seeding procedure without imposing equal seed counts on communities of unequal size. Within the focal community, the most positive initial beliefs are assigned to a fraction $x$ of the highest-degree users (see Appendix~\ref{app:Empirical Networks-bitcoin} for initialization details).

Figure~\ref{fig:bitcoin-otc-with-network}(b) shows how the long-run belief ($b^*$) varies with prestige bias $p_s$, trust--distrust coupling $\rho$, and promoter prevalence $x$. Under \rrptE (\pl{top} row), $b^*$ increases monotonically with $p_s$: stronger prestige bias shifts the long-run belief toward the positively seeded promoters. This shift is stronger when $x$ is larger, with larger promoter prevalence producing a steeper increase in $b^*$ as $p_s$ rises. For a fixed $x$, weaker trust--distrust coupling ($\rho\rightarrow 0$) produces a more positive long-run belief than tighter inverse coupling ($\rho$ closer to $-1$), because the retained high-trust relationships are more selectively directed toward hubs under prestige bias.

Under \rrptF, the trend reverses (\pl{bottom} row). With $\rho$ fixed, $b^*$ decreases as $p_s$ increases. When $x$ is fixed, the response to prestige bias depends on trust--distrust coupling: for the nearly perfectly inversely coupled case $(\rho\rightarrow-1)$, the final belief remains positive and largely insensitive to $p_s$; for weaker coupling, $b^*$ becomes negative and decreases further as $p_s$ rises. 

On a reputation network with trust anchored in observed ratings, the Bitcoin-OTC results follow the same regime-dependent pattern as the synthetic networks: prestige bias raises the long-run contribution of positively seeded hubs under \rgE filtering and lowers it under \rgF filtering. Because distrust is generated through the same trust--distrust coupling used in the synthetic model, the Bitcoin-OTC analysis provides a semi-empirical structural application. It examines whether the regime-dependent pattern persists when one relational dimension is empirically grounded in observed evaluations.

\paragraph{Dynamics on the U.S.\ state legislators' follow network.}

The legislator graph is a directed follow network on Twitter/X among U.S. state legislators, with {\bf 3{,}152} legislators and {\bf 107{,}785} directed follow ties (Figure~\ref{fig:trend-legislative-with-network}(a)). Party affiliation defines two communities: blue vertices denote Democratic legislators, and red vertices denote Republican legislators. 
This case provides an empirical network evaluation in which the two relational dimensions are constructed from distinct observed signals. Structural proximity in the follower network serves as an operational proxy for relational affinity, whereas ideological divergence in audience composition serves as an operational proxy for relational skepticism or trust--distrust conflict. These data-derived quantities provide empirically grounded inputs to the gated model and allow us to assess whether the regime-dependent structural mechanism persists when trust and distrust are operationalized from observed network and ideological signals rather than generated through the controlled coupling process used in the synthetic and Bitcoin-OTC analyses. Within the two partisan communities, there are noticeable smaller clusters associated with state-level geographic proximity. 
Each legislator's initial belief is initialized from the Shor--McCarty ideology score (see \methods). As in the Bitcoin-OTC analysis, promoters are seeded in the larger community, which in this network is the Democratic community. On the original Shor--McCarty scale, Democratic legislators have negative scores, whereas the synthetic and Bitcoin-OTC analyses define the seeded community as positively oriented. We therefore reverse the sign of the ideology scores so that the larger seeded community has positive initial beliefs, maintaining a common orientation across analyses. Within the Democratic community, the most positive initial beliefs are assigned to a fraction $x$ of high-degree legislators.

Unlike the synthetic and Bitcoin-OTC networks, the regional clustering structure of this network often prevents global consensus from emerging, particularly under the \rgF regime. To facilitate comparison between regimes, Figures~\ref{fig:trend-legislative-with-network}(b) and (c) show the final network-average belief $\langle b^* \rangle$ 
versus prestige bias $p_s$. In the \rgE regime (Figure~\ref{fig:trend-legislative-with-network}(b)), $\langle b^* \rangle$ rises monotonically with $p_s$ for all $x$ examined; larger $x$ produces a steeper increase as $p_s$ rises.In the \rgF regime (Figure~\ref{fig:trend-legislative-with-network}(c)), $\langle b^* \rangle$ remains insensitive to the prestige bias, with its final value remaining negative as $p_s$ increases.

To understand which actors drive these outcomes, we examine the long-run influence vector $\bar{\gamma}$, the left null vector associated with the directed Laplacian of the filtered influence graph. Each entry $\bar{\gamma}_i$ measures how much legislator $i$'s initial belief contributes to the final belief state. Figure~\ref{fig:trend-legislative-with-network}(d) plots binned summaries of $\bar{\gamma}$ in the \rgE regime versus legislators' pre-filtering in-degree, shown separately for the two communities. Dashed lines represent $P(\bar{\gamma}_i>0)$, the fraction of legislators with positive long-run influence. Solid lines represent $\mathrm{Median}(\bar{\gamma}_i \mid \bar{\gamma}_i>0)$, the median influence among legislators with positive long-run influence. These summaries show that, under the \rgE regime, long-run influence is concentrated among higher-degree legislators.

Under the \rgF regime, the effective influence graph fragments into approximately 40 isolated strongly connected components (SCCs), many of them singletons. Because these components receive no outside influence, their members retain component-level limiting beliefs. As shown in Figure~\ref{fig:trend-legislative-with-network}(e), these SCCs are predominantly occupied by low-degree Republican legislators, even though filtering generally reduces in-degree across the network. In this setting, the role of peripheral legislators arises through structural insulation and persistence rather than broad downstream reach across the full network. Their negative initial beliefs remain represented in the terminal state of their components and contribute negatively to the network-average terminal belief $\langle b^\ast \rangle$. Thus, the legislator case demonstrates that uncertainty-based filtering can alter which initial beliefs remain represented in aggregate terminal outcomes, including through fragmentation of the effective influence graph.

Across the empirical network evaluations, conflict-based filtering alters the structural channels through which initial beliefs contribute to limiting beliefs. In Bitcoin-OTC, the modeled filtering mechanism changes the relative direct long-run contribution of positively seeded hubs under \rgE and \rgF conditions. In the legislator network, \rgF filtering additionally creates fragmented components in which peripheral actors retain component-level limiting beliefs through structural insulation.

\section{Discussion}

This work shows that collective belief dynamics can depend on how ambivalent relational evaluations determine which influence pathways remain active. Under the specified model configuration---degree-aligned trust, degree-biased positive seeding, trust--distrust coupling, and dual-threshold gating---the same initial placement of positive beliefs among high-degree agents produces opposite outcomes across regimes. \rgE filtering preferentially retains influence pathways directed toward high-degree agents, increasing their direct long-run contribution to the limiting belief. In contrast, \rgF filtering can render many hub-directed pathways inactive in the effective influence graph when their uncertainty exceeds the admission threshold, reducing the direct long-run contribution of positively seeded hubs and shifting the limiting belief away from the seeded direction. The matched one-dimensional comparison in Appendix~\ref{app:trend_1D_case} provides a supporting diagnostic: under the tested net-trust thresholds, filtering on net trust alone does not produce the observed reversal. This result is consistent with the interpretation that the uncertainty coordinate enables the contrasting \rgE and \rgF outcomes reported in the main analyses.

The results also show that the consequences of conflict-sensitive filtering depend on the connectivity of the resulting effective influence graph. When the graph remains connected, as in the consensus settings examined in the synthetic and Bitcoin-OTC analyses, filtering changes which agents' initial beliefs directly contribute to the shared limiting belief. When filtering fragments the graph, as in the legislator network, peripheral source components can retain component-level limiting beliefs, allowing their initial beliefs to remain represented in the network-average terminal outcome despite limited broader reach. This possibility is relevant for polarized or modular networks, where conflict-sensitive admission can structurally insulate subgroups and preserve locally concentrated belief positions rather than allowing them to be integrated through network-wide influence.

\paragraph{The limits of centrality.}
High-degree and centrally embedded actors are often treated as privileged sites of diffusion and social influence \cite{kitsak2010,centola2010}. This interpretation assumes that structural centrality is a reliable proxy for influence. Our results show that this relationship can weaken when social interactions are filtered through relational evaluations of trust and distrust. A source may begin with high structural prominence yet lose asymptotic weight when filtering removes the relationships through which its beliefs would otherwise propagate.

This distinction is relevant in contemporary epistemic environments, where public institutions, media organizations, political elites, and platform influencers can combine high visibility with contested credibility \cite{cologna2025,lazer2018}. A source may be widely recognized yet fail to shape belief if audiences experience it as conflicted, self-interested, or institutionally distrusted. The model therefore extends work on networked information processing by showing how relational ambivalence can reshape the effective pathways through which beliefs propagate \cite{finkel2020,lorenzspreen2023}.

\paragraph{From targeting actors to diagnosing filtering regimes.} 
The framework suggests that analyses of influence interventions should first identify which part of the model an intervention is expected to affect. An intervention may alter relational evaluations, such as trust, distrust, or perceived conflict; it may change the thresholds through which receivers admit influence; or it may modify the network of available exposure. The present simulations vary the thresholds and relational inputs as analytical parameters, but they do not estimate the effects of real-world interventions. Within the model, interventions that increase trust or reduce trust--distrust conflict make more relationships admissible, whereas interventions that raise conflict or lower net trust would remove influence channels. Changes in thresholds instead alter the community's general tolerance for net trust and relational conflict.  This connects the model to research showing that misinformation response depends on credibility, attention, and the social setting in which information is received \cite{ecker2022,pennycook2021}.

\paragraph{Ambivalence as an empirical object.}
The model also identifies a measurement challenge for empirical network science. Signed-network datasets often encode relationships as positive or negative, which is useful for studying balance, trust prediction, and signed consensus but can compress relationships that differ in the joint levels of trust and distrust. The two empirical network evaluations illustrate complementary ways to preserve this information: Bitcoin-OTC combines observed graded trust ratings with a modeled distrust dimension, whereas the legislator network constructs relational affinity and conflict from distinct observed signals. Future datasets should measure such dimensions separately, including trust and distrust, confidence and suspicion, or expertise and motive.

\paragraph{Scope and future directions.}
The present results concern one parsimonious conflict coordinate, a family of trust--distrust couplings ranging from strongly inverse to weak, degree-based prestige allocation, and additive-exposure dynamics. Alternative conflict functions, positive trust--distrust coupling, heterogeneous filtering thresholds, and fixed-attention-budget updating may produce different quantitative boundaries or qualitative regimes. Future work should examine which aspects of the hub--periphery reversal persist across these alternatives, including settings in which credibility is shaped by group memberships or higher-order social contexts \cite{battiston2020,iacopini2019}. It should also consider heterogeneous agents and multidimensional beliefs, because different individuals may tolerate trust--distrust conflict differently and may evaluate the same source differently across topics.

This paper shows that network position does not determine long-run influence independently of relational evaluation. When trust and distrust coexist, the criteria that determine which relationships remain admissible can change the degree profile of retained pathways and agents' direct contributions to limiting beliefs. In fragmented networks, peripheral components may retain their limiting beliefs through structural insulation and thereby affect the network-wide average limiting belief. Ambivalence is therefore a mechanism through which communities determine who is permitted to shape belief.

\section{Methods and Materials}

\subsection{Gated Network Credence Belief Dynamics}\label{sec:belief-model}

We consider a society of $N$ agents connected by a fixed directed social network with adjacency matrix $A \in \{0,1\}^{N \times N}$, where $A_{ij}=1$ if agent $i$ regards agent $j$ as a potential source of influence and $A_{ij}=0$ otherwise. Throughout, the row index denotes the receiver of influence and the column index denotes the source. Each agent $i$ holds a continuous belief $b_i(t) \in \mathbb{R}$.

For each directed relationship, the trust--distrust pair $(\tau_{ij},\delta_{ij})$ is mapped to net trust and uncertainty as
\[
T_{ij}=\tau_{ij}-\delta_{ij}, \qquad U_{ij}=\tau_{ij}+\delta_{ij}-1.
\]
Net trust captures whether the receiver is inclined to rely on the source, while uncertainty captures the degree of internal conflict between trust and distrust in the same relationship. Under the dual-threshold rule of Equation~\eqref{eq:decision}, influence from $j$ to $i$ is admitted only when both conditions are simultaneously satisfied. Retaining only these admissible ties defines the masked adjacency matrix $A^{\mathrm{IN}}$:
\[
A^{\mathrm{IN}}_{ij} =
\begin{cases}
1, & \text{if } A_{ij}=1 \ \text{and} \ T_{ij}\ge T_T \ \text{and} \ 
     U_{ij}\le U_T,\\
0, & \text{otherwise.}
\end{cases}
\]
Beliefs then evolve as
\begin{align}\label{eq:dynamics}
\frac{db_i(t)}{dt} &= \sum_j A^{\mathrm{IN}}_{ij}\bigl(b_j(t)-b_i(t)\bigr) \notag\\
                   &= \sum_j A^{\mathrm{IN}}_{ij}\,b_j(t)
                     - b_i(t)\sum_j A^{\mathrm{IN}}_{ij}.
\end{align}
This equation states that agent $i$ is pulled toward the beliefs of the sources that survive its trust--uncertainty gate. We use an unnormalized continuous-time Laplacian formulation: the diagonal
term below is the self-decay term that balances the total influence entering from admitted sources. 
Each admitted tie enters the update with the same additive coefficient. Trust and distrust therefore determine which influence channels remain active, while the binary mask $A^{\mathrm{IN}}$ determines the effective topology and the number of admitted sources contributing to each receiver's update.

Letting
\[
D^{\mathrm{row}}=\operatorname{diag}\!\bigl(k^{\mathrm{row}}_1,\ldots,k^{\mathrm{row}}_N\bigr),
\qquad
k_i^{\mathrm{row}}=\sum_j A^{\mathrm{IN}}_{ij},
\]
denote the diagonal matrix of admitted-source counts for each receiver, Equation~\eqref{eq:dynamics} takes the compact form
\begin{equation}\label{eq:laplacian_dynamics}
\dot{b}(t)=-L\,b(t), \qquad L=D^{\mathrm{row}}-A^{\mathrm{IN}}.
\end{equation}
Unlike the Laplacian of an undirected graph, here $L$ is generally non-Hermitian, so its eigenvectors do not form a complete basis of orthonormal eigenvectors, and $L$ may not be diagonalizable. Nevertheless, the left and right null spaces of $L$ determine the long-run belief dynamics, identifying which parts of the admissible-influence network remain effective in shaping the final collective belief~\cite{veerman2019}.

The next subsection describes this spectral structure and shows how the trust--distrust filtered graph determines the consensus, fragmentation, and hub--periphery patterns reported in the Results.

\subsection{Spectral characterization of long-run beliefs}\label{sec:null-space}

The Results section uses long-run influence weights to explain why hubs dominate in the \rgE regime but lose influence in the \rgF regime. These weights are obtained from the null space of the filtered Laplacian. Once the trust--distrust gate fixes $A^{\mathrm{IN}}$, the directed Laplacian $L$ determines which initial beliefs persist, which agents inherit those beliefs, and whether the network reaches consensus or fragments into multiple limiting belief components. The left null space identifies the agents whose initial beliefs influence the limiting belief values, while the right null space determines how those limiting belief values are distributed across the network.

The graph induced by $A^{\mathrm{IN}}$ contains $K$ reaches: maximal subsets of nodes $R_1,\ldots,R_K$ such that each $R_m$ is the set of all nodes reachable from some root, and no strictly larger such set exists. Reachability is defined in the influence direction: source $j$ points to receiver $i$ whenever $A^{\mathrm{IN}}_{ij}=1$. Within each reach $R_m$ there is a unique cabal $B_m$: a source strongly connected component in the influence direction that receives no influence from outside itself and from which every node in $R_m$ is reachable. The long-run belief vector is then
\begin{equation}\label{eq:longrun}
\lim_{t\to\infty} b(t)=\sum_{m=1}^{K}\bigl(\bar{\gamma}_m^\top b(0)\bigr)\gamma_m,
\end{equation}
where $\bar{\gamma}_m$ is the left null vector of $L$ associated with $R_m$, normalized so that its entries are nonnegative, sum to one, and are strictly positive only on the cabal $B_m$~\cite{veerman2019}. Consequently, only the initial beliefs of cabal nodes contribute directly to the composition of the long-run belief components. The entries of $\bar{\gamma}_m$ provide an asymptotic influence measure: larger entries indicate that an agent's initial belief makes a larger direct contribution to the limiting belief state. This is the quantity used in the Results to diagnose whether influence is concentrated on hubs or redistributed toward lower-degree agents after filtering.
In our analyses, we distinguish four related measures that characterize different aspects of influence after filtering. Link retention refers to the number of incoming relationships that survive the gate. Direct long-run contribution refers to the coefficient $\bar{\gamma}_i$ multiplying agent $i$'s initial belief in a component-level limiting belief. Persistence refers to the retention of an initial belief within an isolated or source component. Broad downstream influence refers to the capacity of an agent or component to affect a nontrivial reachable set of other nodes. These quantities are related through the filtered graph, yet they need not coincide when the graph contains multiple reaches or isolated strongly connected components. Accordingly, post-filtering in-degree diagnoses retained connectivity, whereas $\bar{\gamma}_i$ measures direct long-run contribution. The downstream extent of that contribution additionally depends on the size and accessibility of the associated reach.

Global consensus arises if and only if the filtered graph has exactly one reach ($K=1$); in this case the asymptotic influence vector $\bar{\gamma}$ reduces to the unique cabal-supported left null vector $\bar{\gamma}_1$. When $K>1$, different reaches contribute separate limiting belief components, corresponding to the fragmentation outcomes shown in the threshold phase diagrams. Thus, the same spectral representation links the consensus--fragmentation transition to the hub--periphery reversal: the thresholds determine the filtered graph, the filtered graph determines the cabals and null vectors, and the null vectors determine which agents shape the long-run belief state.

\paragraph{\textbf{Modeling note: additive social pull.}}

We use the unnormalized Laplacian
$L=D^{\mathrm{row}}-A^{\mathrm{IN}}$
because the number of admitted relationships is part of the mechanism under study. Each surviving tie contributes an additive unit of social pull. Filtering therefore changes both the composition of admissible sources and the total interpersonal input received by a focal agent.
This formulation represents settings in which a relationship becoming unacceptable or highly conflicted reduces effective social exposure rather than triggering immediate reallocation of attention to other available sources. Under this assumption, a receiver with fewer admitted sources experiences less total social pull.
A row-normalized alternative,
$L_{\mathrm{norm}}=I-\left(D^{\mathrm{row}}\right)^{-1}A^{\mathrm{IN}},$
encodes a fixed attention-budget assumption: each receiver redistributes a constant total influence across sources that remain admissible. The latter normalization form describes a different behavioral setting. In this study, we concern additive-exposure dynamics, in which filtering changes the intensity as well as the composition of social influence.

\subsection{Experimental Design and Parameterization}\label{sec:simulation-setup}

The simulations operationalize the mechanisms introduced in the Introduction and evaluated in the Results:
ambivalent trust--distrust ties, dual trust--distrust filtering, prestige-biased trust
allocation, positive extreme seeding on prominent agents, and community-level polarization.
We run a parameterized experiment on three synthetic network topologies: \nameER (ER) random networks, \nameBA (BA) scale-free networks, and modular networks with explicit community structure. We then repeat the analysis on two empirical networks. The synthetic networks isolate how the mechanism behaves under random mixing, degree heterogeneity, and community structure, while the empirical networks test whether the same regime-dependent reversal appears in observed social systems.

The four experimental parameters are designed as stylized counterparts of empirical features that often co-occur in social information environments. At the edge level, the trust--distrust coupling $\rho$ captures whether trust and distrust operate as opposite ends of a single scale or as separable evaluations that can coexist within the same relationship, consistent with prior work on relational ambivalence~\cite{lewicki1998,moody2014}. The prestige bias $p_s$ controls how strongly trust is aligned with source prominence, reflecting the role of visibility, status, and network position in social influence and diffusion~\cite{kitsak2010,centola2010}. At the node level, the promoter prevalence $x$ controls how widely extreme positive initial beliefs are assigned among more highly connected actors, allowing us to test whether seeded prominence is amplified or suppressed under different filtering regimes.
In modular networks, the polarization parameter $\psi$ controls the initial separation between communities, reflecting the role of homophily, partisan sorting, and community structure in polarized information environments. These four experiment parameters are used to assess how empirically motivated forms of ambivalence, prominence, concentrated seeding, and community separation interact to determine whether hubs are amplified or sidelined.

\subsubsection{Trust--Distrust Coupling}\label{app:trust-distrust-coupling-definition}

To model relationships in which distrust ranges from an almost opposite evaluation of trust to a weakly coupled evaluation, we specify:
\begin{equation}
\delta_{ij}=\rho\left(\tau_{ij}-\frac{1}{2}\right)+\frac{1}{2}+\epsilon,
\label{eq:trust_distrust_coupling}
\end{equation}
where $\rho \in [-1,0]$ controls the strength of the inverse-to-weak coupling between trust and distrust, and $\epsilon \sim \mathcal{N}(0,\sigma_{\epsilon}^{2})$ is a zero-mean Gaussian noise term. At $\rho=-1$, trust and distrust approximate a one-dimensional opposition: higher trust is systematically paired with lower distrust. As $\rho$ approaches $0$, this association weakens, allowing trust and distrust to coexist more readily within individual relationships because residual variation in $\epsilon$ becomes more consequential.

For each directed edge $(i,j)$, we first sample $\tau_{ij} \sim U(0,1)$. We then sample $\epsilon$ and compute $\delta_{ij}$ from Equation~\eqref{eq:trust_distrust_coupling}. Because $\delta_{ij}$ is required to remain in $[0,1]$, any draw of $\epsilon$ that yields $\delta_{ij} \notin [0,1]$ is rejected and resampled until a valid value is obtained.

The present parameterization focuses on inverse and weakly coupled trust--distrust relationships. Positive population-level coupling, under which highly trusted sources also tend to receive high distrust across relationships, represents a distinct joint-evaluation structure and remains a useful direction for future extensions.

\subsubsection{Prestige-Biased Trust Assignment}\label{prestige-bias-rule}

Source prominence is measured by the pre-filtering degree of the potential source node. In the directed network convention used here, this is the degree of node $j$ as a source of influence before the trust--distrust mask is applied. Node degree $k_j$ is used as a proxy for perceived prestige, so that neighbors with larger degree are treated as more prestigious. For each agent $i$, let $\mathcal{N}(i)$ denote the set of potential sources of $i$. The values $\{(\tau_{ij},\delta_{ij})\}_{j\in\mathcal{N}(i)}$ are first generated according to Equation~\eqref{eq:trust_distrust_coupling}.

To make the assignment reproducible while preserving the marginal trust and distrust distributions, trust--distrust pairs are then permuted across the potential sources of each receiver. Specifically, sources are ranked by a prestige-biased score
\[
s_{ij}=p_s\,\widetilde{k}_j+(1-p_s)\eta_{ij},
\]
where $\widetilde{k}_j$ is the normalized pre-filtering degree of source $j$ among $i$'s potential sources and $\eta_{ij}$ is an independent random score. The sampled trust--distrust pairs are sorted by $\tau_{ij}$ and assigned in descending order to sources sorted by $s_{ij}$. Because this procedure only permutes the existing $\tau$ and $\delta$ values across $\mathcal{N}(i)$, their marginal distributions are preserved. 
This procedure creates a controlled contrast between degree-independent trust allocation and degree-aligned trust allocation, while preserving the receiver-specific marginal distributions of sampled trust--distrust pairs. It operationalizes prestige bias as a rank association between source prominence and received trust rather than as a claim that degree universally measures prestige or that all communities trust prominent sources more strongly.
When $p_s=0$, assignment is random with respect to source degree; when $p_s=1$, trust is maximally aligned with source prestige. Intermediate values of $p_s$ generate partial rank alignment between trust and source prominence.

\subsubsection{Promoter Prevalence: Degree-Biased Positive Seeding}\label{subsec:promoter-prevalence-mech}

To control promoter prevalence, we introduce a parameter $x \in [0,1]$, representing the fraction of highest-degree nodes assigned the most positive initial beliefs. This degree-biased seeding serves as a counterfactual test of how a concentration of positive extreme beliefs among prominent sources interacts with the filtering regime. Initial beliefs are first sampled independently as $b_i(0) \sim U(-B_{\max},B_{\max})$.

Let $\mathcal{H}_x$ be the set of nodes in the top fraction $x$ of the degree ranking, and let $\mathcal{P}_x$ be the set of nodes in the top fraction $x$ ranked by initial belief value $b_i(0)$. Thus, $\mathcal{P}_x$ contains the most positive initial beliefs, rather than the largest absolute-magnitude beliefs. We sort $\mathcal{H}_x$ in descending order by degree and $\mathcal{P}_x$ in descending order by belief value. Beliefs are reassigned pairwise so that the highest-degree node in $\mathcal{H}_x$ receives the most positive belief in $\mathcal{P}_x$, the second-highest-degree node receives the second-most positive belief, and so on. Equivalently, if a high-degree node does not already hold the corresponding positive extreme belief, its belief is swapped with that of the matched node in $\mathcal{P}_x$. This preserves the overall belief distribution while concentrating the most positive initial beliefs among high-degree nodes, with positivity increasing with degree.

\subsubsection{Inter-Community Polarization}\label{polarization}

To introduce inter-community polarization while preserving the marginal distribution of initial beliefs, we consider a modular network consisting of two equally sized communities, $C_1$ and $C_2$, with $|C_1|=|C_2|=N/2$. Each agent is initially assigned a belief independently according to
$b_i(0)\sim\mathcal{U}(-B_{\max},B_{\max})$,
independently of community membership. We designate $C_1$ as the positive pole and $C_2$ as the negative pole.

For a belief configuration $\mathbf{b}$, we define the degree of inter-community polarization as
\[
P(\mathbf{b})
=
\frac{2}{N}
\left(
\sum_{i\in C_1} b_i
-
\sum_{j\in C_2} b_j
\right).
\]
Given the initial belief vector $\mathbf{b}^{(0)}$, we construct a maximally polarized configuration $\mathbf{b}^{\max}$ by sorting the $N$ belief values and assigning the $N/2$ largest values to $C_1$ and the $N/2$ smallest values to $C_2$. Thus,
\[
P_{\max}=P(\mathbf{b}^{\max})
\]
is the maximum polarization attainable by reassigning the existing belief values across the two communities. This construction preserves the overall empirical distribution of beliefs, as it only changes their assignment to agents.

Let
\[
P_0=P(\mathbf{b}^{(0)})
\]
denote the polarization of the initial assignment. For a prescribed polarization parameter $\psi\in[0,1]$, we set the target polarization to
\[
P_{\mathrm{target}}
=
P_0+\psi\left(P_{\max}-P_0\right).
\]
Thus, $\psi$ parametrizes the fraction of the available polarization induced relative to the initial configuration: $\psi=0$ leaves the initial assignment unchanged, whereas $\psi=1$ sets the target at the maximally polarized configuration.

To reach this target, we repeatedly sample a pair $(i,j)$ uniformly from $C_1\times C_2$. If
\[
b_i<b_j,
\qquad i\in C_1,\;j\in C_2,
\]
the two beliefs are exchanged. This operation moves the larger belief toward the positive community $C_1$ and the smaller belief toward the negative community $C_2$. For an accepted swap, the change in polarization is
\[
P(\mathbf{b}')-P(\mathbf{b})
=
\frac{4}{N}(b_j-b_i)>0.
\]
The procedure is repeated until the polarization reaches or exceeds $P_{\mathrm{target}}$. Because the procedure only exchanges existing belief values, it preserves the global distribution of initial beliefs while changing their allocation across communities.

\subsubsection{Experimental Factors and Parameter Ranges}

The parameter ranges are chosen to span substantively distinct cases of these mechanisms: nearly one-dimensional versus weakly coupled trust--distrust evaluations, low versus high prestige bias, sparse versus widespread assignment of positive promoters among more highly connected agents, and low versus high inter-community belief separation. Holding the noise level fixed at $\sigma_\epsilon=0.05$, we vary
\[
\rho\in\{-1,\,-0.7,\,-0.4,\,-0.01\},\qquad
p_s\in\{0.05,\,0.10,\,0.40,\,0.60,\,0.80,\,1\}.
\]
Agent beliefs are initialized independently as
\[
b_i(0)\sim\mathcal{U}(-B_{\max},B_{\max}), \qquad B_{\max}=5.
\]
To vary the prevalence of positive promoters among high-degree nodes, we use
\[
x\in\{0.05,\,0.10,\,0.15,\,0.20\}.
\]
In modular networks, positive promoters are seeded within one community so that the simulations test whether regime-dependent influence remains localized, crosses communities, or is counteracted by the opposite community. We examine inter-community polarization in Appendix~\ref{app:polarization-role-modular-synthetic} using $\psi\in\{0,\,0.4,\,1\}$, corresponding to unpolarized, intermediate, and strongly polarized initial belief configurations. Prestige bias is applied within modules in the modular simulations, reflecting the assumption that prestige is evaluated primarily within community boundaries rather than across weak inter-community ties.
\subsubsection{Network Topologies and Empirical Data}\label{Networks}

\paragraph{Synthetic networks.}
The synthetic networks are selected to isolate three structural conditions
relevant to the hub--periphery reversal: random mixing, heterogeneous degree,
and community structure. The three synthetic topologies have $N=5000$ nodes each. The \nameER (ER) network is generated as an undirected graph in which each pair of nodes is connected independently with probability $p=0.01$, providing a random-mixing baseline with relatively homogeneous degree. The \nameBA (BA) network is grown by preferential attachment with $m=25$ links per arriving node, producing a heavy-tailed degree distribution in which hub effects are expected to be strongest. The modular network consists of two equal-size communities ($N_1=N_2=2500$), each generated independently as a BA network with $m_1=m_2=21$, followed by the addition of randomly selected cross-community edges to obtain modularity approximately $Q=0.33$.
This topology tests whether the regime-dependent reversal persists when influence is constrained by community boundaries.

\paragraph{Directed relationships.}
Each synthetic topology is first generated as an undirected graph. Every undirected edge $\{i,j\}$ is then represented by two potential directed relationships, $(i,j)$ and $(j,i)$. The two directions receive independently sampled trust--distrust values and are admitted separately by the dual threshold rule. Thus, the underlying potential graph is bidirectional, but the effective influence graph $A^{\mathrm{IN}}$ is generally asymmetric. Directionality therefore enters through the relational evaluations and the gate, which produces the non-Hermitian directed Laplacian used in the dynamics.

\paragraph{Replication.}
For each synthetic network topology and parameter combination, we generate 100
independent realizations. Each realization redraws the network, initial beliefs,
trust--distrust values, and the prestige-based assignment of relational
evaluations. Thus, the synthetic results average over 100 independently
generated networks per topology. The same 100 networks and initial belief
configurations are reused across parameter combinations, so that differences between parameter values are not confounded by network-to-network variability. For the empirical networks, the observed topology is held fixed; repeated runs redraw only the quantities generated or randomized within the model, including distrust in Bitcoin-OTC, initial beliefs, and any randomized assignment or initialization procedure. Error bars denote $\pm 1$ standard deviation across the corresponding 100 runs.

\paragraph{Empirical networks.}
The empirical analyses test the same mechanism in two observed settings with different sources of relational evaluation. The Bitcoin-OTC reputation graph ($N=5{,}881$ users, $35{,}592$ directed ratings) provides a directed online marketplace network in which users assign graded reputational scores to one another. These ratings preserve more information than binary positive/negative ties and allow trust values to be anchored in observed evaluations. For each directed rating $i\to j$, the empirical score $\mathrm{Sign}_{ij}\in[-10,10]$ is rescaled to
\[
m_{ij}=\frac{\mathrm{Sign}_{ij}+10}{20}\in[0,1].
\]
We use $m_{ij}$ to construct the empirical trust value for that directed relationship. Because independent distrust is not directly observed in this dataset, distrust is generated using the same controlled trust--distrust coupling in Equation~\eqref{eq:trust_distrust_coupling}; this allows us to examine how varying ambivalence changes influence on a real reputation network while keeping trust anchored in observed ratings.

The second empirical system is a directed follow network of U.S. state legislators on X ($N=3{,}152$ accounts, $107{,}785$ follows). This network provides a polarized institutional communication setting in which structural affinity, partisan organization, and ideological separation can be measured separately. It is therefore useful for testing whether the hub--periphery reversal persists when trust and distrust are inferred from distinct observed signals rather than generated through a synthetic coupling process.

For the legislator network, the trust and distrust dimensions are operationalized from distinct observed relational and ideological signals. We represent trust by structural affinity in the broader follower network, measured as the cosine similarity between node2vec embeddings learned from the full directed follower graph~\cite{grover2016node2vec}; only the embeddings corresponding to legislators are subsequently retained for constructing trust values on legislator-to-legislator model ties. We represent distrust by ideological divergence in audience composition, measured as the normalized absolute difference between legislators' median follower-audience leaning scores. Thus, the legislator-to-legislator follow network defines the topology of the empirical opinion-dynamics model, whereas the broader follower network contributes to the construction of the empirical relational proxies.

These measures serve as observed-signal inputs to the model rather than direct psychological measurements of legislator-to-legislator trust and distrust. We assess their construct consistency and behavioral correspondence using Shor--McCarty ideological distance, retweet behavior, and sentiment in reply and mention networks, with dyadic network dependence addressed using QAP permutation tests. Full construction details and QAP statistics are provided in Appendix~\ref{app:legislative network}.

\FloatBarrier
\section*{Supplementary information}
Sections, figures and tables numbered with an ``S'' prefix refer to the Supplementary Information, which is
distributed with this preprint as an ancillary file
(\texttt{anc/Supplementary\_Information.pdf}) and can be downloaded from the
``Ancillary files'' link on the arXiv abstract page.

\section*{Data availability}
The Bitcoin-OTC reputation network analyzed in this study is publicly
available~\cite{kumar2016}. The U.S.\ state-legislator follow network was
constructed from the political-elite dataset described in prior
work~\cite{biswas2025political,biswas2026attention}; raw platform data are not
redistributed. The processed network data, the derived trust/distrust attributes,
and the preprocessing notebooks that regenerate them from the source files are
archived at Zenodo, \url{https://doi.org/10.5281/zenodo.22063434}~\cite{masoumi2026gnccode},
and are also available at
\url{https://github.com/RahaMasoumi/Gated-Network-Credence}.

\section*{Code availability}
The custom code that produces the main findings of this study is archived at
Zenodo, \url{https://doi.org/10.5281/zenodo.22063434}~\cite{masoumi2026gnccode},
and is available at \url{https://github.com/RahaMasoumi/Gated-Network-Credence}
under the MIT license. The code is written in Python 3.10 and requires
\texttt{numpy}, \texttt{scipy}, \texttt{networkx} and \texttt{matplotlib}.

\section*{Acknowledgements}
The authors would like to acknowledge support from AFOSR, NSF \#2318461, and Pitt Cyber Institute's PCAG awards. The research was partly supported by Pitt's CRCD resources. Any opinions, findings, and conclusions or recommendations expressed in this material do not necessarily reflect the views of the funding sources.

\section*{Author contributions}
RM contributed to the study design and performed the analyses. YRL designed the research and overall study methodology. RM and YRL drafted the manuscript. AB contributed to the analyses of empirical data. All authors revised the manuscript and approved the submitted version.

\section*{Declaration of AI use}
During the preparation of this work, the authors used LLMs to polish or refine the language. The authors then reviewed and edited the content and take full responsibility for the content of the publication.

\section*{Competing interests}
The authors declare no competing interests.

\FloatBarrier
\bibliographystyle{unsrtnat}
\bibliography{references_main}

\end{document}